\documentclass[namedreferences]{kluwer}    

\usepackage{graphicx}

\newcommand{\Hb}{\ensuremath{{\rm H}\beta}}
\newcommand{\Mgb}{\ensuremath{{\rm Mg}\, b}}
\newcommand{\aFe}{\ensuremath{\alpha/{\rm Fe}}}
\newcommand{\ZH}{\ensuremath{Z/{\rm H}}}

\begin{document}

\bibliographystyle{klunamed}

\begin{article}

\begin{opening}         

\title{Kinematics and stellar populations of 17 dwarf early-type
galaxies\thanks{Based on observations obtained at the Calar Alto
Observatory, Spain.}}

\author{Daniel \surname{Thomas}} \author{Ralf
\surname{Bender}} \author{Ulrich \surname{Hopp}} \author{Claudia
\surname{Maraston}}

\institute{MPI f\"ur extraterrestrische Physik,
Giessenbachstra\ss e, 85748 Garching, Germany\\
Universit\"ats-Sternwarte M\"unchen, Scheinerstr.\ 1, 81679 M\"unchen,
Germany}

\author{Laura \surname{Greggio}}

\institute{Osservatorio Astronomico di Padova, Vicolo
dell'Osservatorio, Padova, Italy}

\runningauthor{D.\ Thomas, R.\ Bender, U.\ Hopp, C.\ Maraston, and L.\
Greggio}
\runningtitle{Kinematics and stellar populations of 17 dwarf
early-type galaxies}

\begin{abstract}
We present kinematics and stellar population properties of 17 dwarf
early-type galaxies in the luminosity range $-14\geq M_B\geq -19$.
Our sample fills the gap between the intensively studied giant
elliptical and Local Group dwarf spheroidal galaxies.  The dwarf
ellipticals of the present sample have constant velocity dispersion
profiles within their effective radii and do not show significant
rotation, hence are clearly anisotropic. The dwarf lenticulars,
instead, rotate faster and are, at least partially, supported by
rotation. From optical Lick absorption indices, we derive
metallicities and element abundances.  Combining our sample with
literature data of the Local Group dwarf spheroidals and giant
ellipticals, we find a surprisingly tight linear correlation between
metallicity and luminosity over a wide range: $-8\geq M_B\geq
-22$. The \aFe\ ratios of our dwarf ellipticals are significantly
lower than the ones of giant elliptical galaxies, which is in
agreement with spectroscopy of individual stars in Local Group dwarf
spheroidals. Our results suggest the existence of a clear kinematic
and stellar population dichotomy between dwarf and giant elliptical
galaxies. This result is important for theories of galaxy formation,
because it implies that present-day dwarf ellipticals are not the
fossiled building blocks of giant ellipticals.
\end{abstract}

\end{opening}           

\section{Introduction}  
Although dwarf galaxies are by far more abundant than giant galaxies,
our knowledge of the kinematics and stellar population properties of
these objects is still very poor. In the framework of hierarchical
clustering dwarf galaxies play an important role as they may be the
seeds for the formation of larger galaxies. It is still not clear,
however, whether dwarf elliptical galaxies are related to giant
ellipticals or form a separate family. A continuity with respect to
mean radii and surface brightnesses \cite{Nieto88} are arguments
against, the vast differences in core properties \cite{Ko85}, instead,
support the existence of a dichotomy between dwarf and giant
ellipticals.  Are the present-day dwarf ellipticals the fossiled
building blocks of giant ellipticals as early suggested by
\inlinecite{DS86}?  We aim to address these questions via a detailed
spectroscopic analysis of the internal kinematic structure and the
stellar population properties, i.e.\ metallicities and element
abundance ratios, of dwarf early-type galaxies.

\medskip
In several observing runs (1995 to 2001) we used the TWIN spectrograph
at the 3.5m telescope on Calar Alto Observatory (Spain) to take
long-slit (along the major axis) spectra of 17 dwarf early-type
galaxies in the Virgo cluster and in the field
\cite{BST85,VC94}. Covering the wavelength range
$4800\lsim\lambda\lsim 5400$~\AA, we obtained a spectral resolution
$\sigma\sim 30$ km/s with $S/N\sim 40$~\AA$^{-1}$ (within 1/2
$r_e$). The line-of-sight velocity distributions were determined from
the Mg triplet near $5175$~\AA\ with the Fourier-Correlation-Quotient
method \cite{Bender90a} and carefully tested with Monte Carlo
simulations. From the Lick absorption-line indices \Hb, \Mgb, Fe5270,
and Fe5335 we derive average ages, total metallicities \ZH, and \aFe\
ratios based on the stellar population models of \inlinecite{TMB02b}.

\section{Kinematics}
\begin{figure}
\caption{Velocity dispersion and velocity as functions of radius. The
grey shaded area indicates the maximum amount of rotation for which
the object is still anisotropic, hence $(v/\sigma)^*=\frac{v_{\rm
rot}/\bar{\sigma}}{\sqrt{\epsilon/(1-\epsilon)}}\approx 0.7$.}
\includegraphics[width=0.49\textwidth]{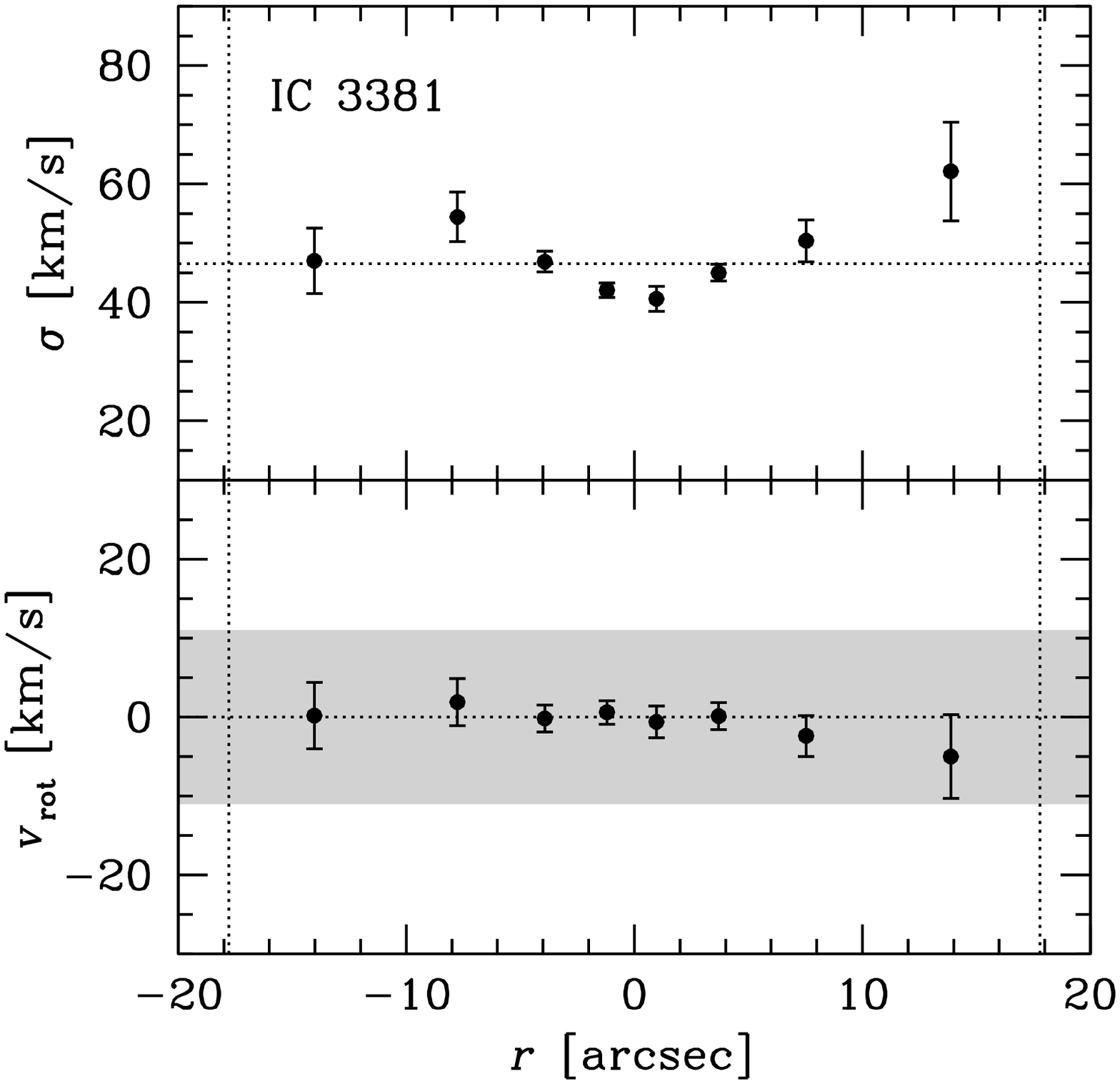}
\includegraphics[width=0.49\textwidth]{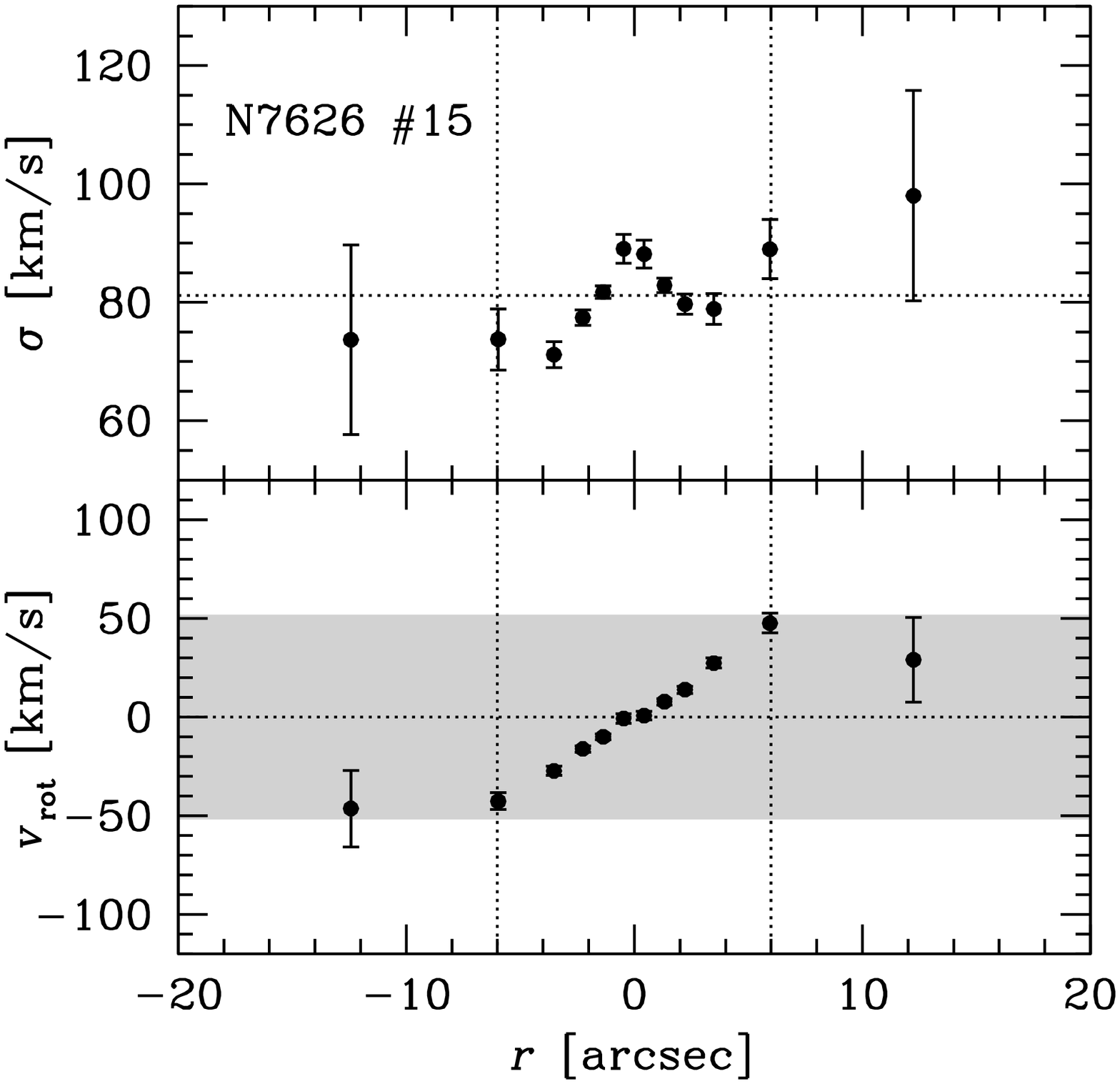}
\end{figure}
In Fig.~1 we show the velocity dispersion ($\sigma$) profiles and
rotation curves ($v_{\rm rot}$) of the Virgo dwarf elliptical IC~3381
(well consistent with \opencite{SiPr02}) and the dwarf lenticular
NGC~7626\#15. All our 14 dwarf ellipticals have profiles similar to
IC~3381 and do not show significant rotation.  We therefore confirm
for a significantly larger sample the early result of
\inlinecite{BN90} and \inlinecite{BPN91}, that dwarf elliptical
galaxies are anisotropic. The 3 dwarf lenticulars, instead, show clear
signs of rotation (see NGC~7626\#15 in Fig.~1). Note that, still, the
latter are not fully flattened by rotation, as the anisotropy
parameter \cite{Ko82} is $(v/\sigma)^*\approx 0.7$ \cite{BBF92}.

These results are in good agreement with the recent studies by
\inlinecite{GGvM02} and Zeilinger et al.\ (this
volume). \inlinecite{Pedetal02}, instead, claim to detect fast
rotation in dwarf ellipticals. We note that the three fast rotators in
their sample are dwarf lenticulars (in agreement with the present
result), and the remaining three dwarf ellipticals have only
negligible rotational velocities and are not rotationally
flattened. The absence of significant rotation sets the clear
distinction from 'normal' low-luminosity ellipticals. Dwarf and giant
elliptical galaxies apparently form separate families.  Note that the
objects for which \inlinecite{SiPr02} (see also this volume) detect
significant rotation are more lumimous and also have higher surface
brightnesses, hence may be at the transition to 'normal'
low-luminosity ellipticals.

\section{Element abundances}
\begin{figure}
\caption{Total metallicity and element abundance ratio \aFe\ as
functions of absolute blue luminosity. Measurements are within 1/2
$r_e$. Filled squares are the dwarf early-type galaxies of this
work. Open squares are dwarf galaxies from Gorgas et al.\ (1997). Open
circles are Local Group dwarf spheroidals from the compilation of
Mateo (1998). Giant ellipticals (filled circles) are from Mehlert et
al.\ (2000) and Beuing et al.\ (2002). The solid line in the left
panel is $[\ZH]=-3.6 - 0.19\cdot M_B$.}
\includegraphics[width=0.49\textwidth]{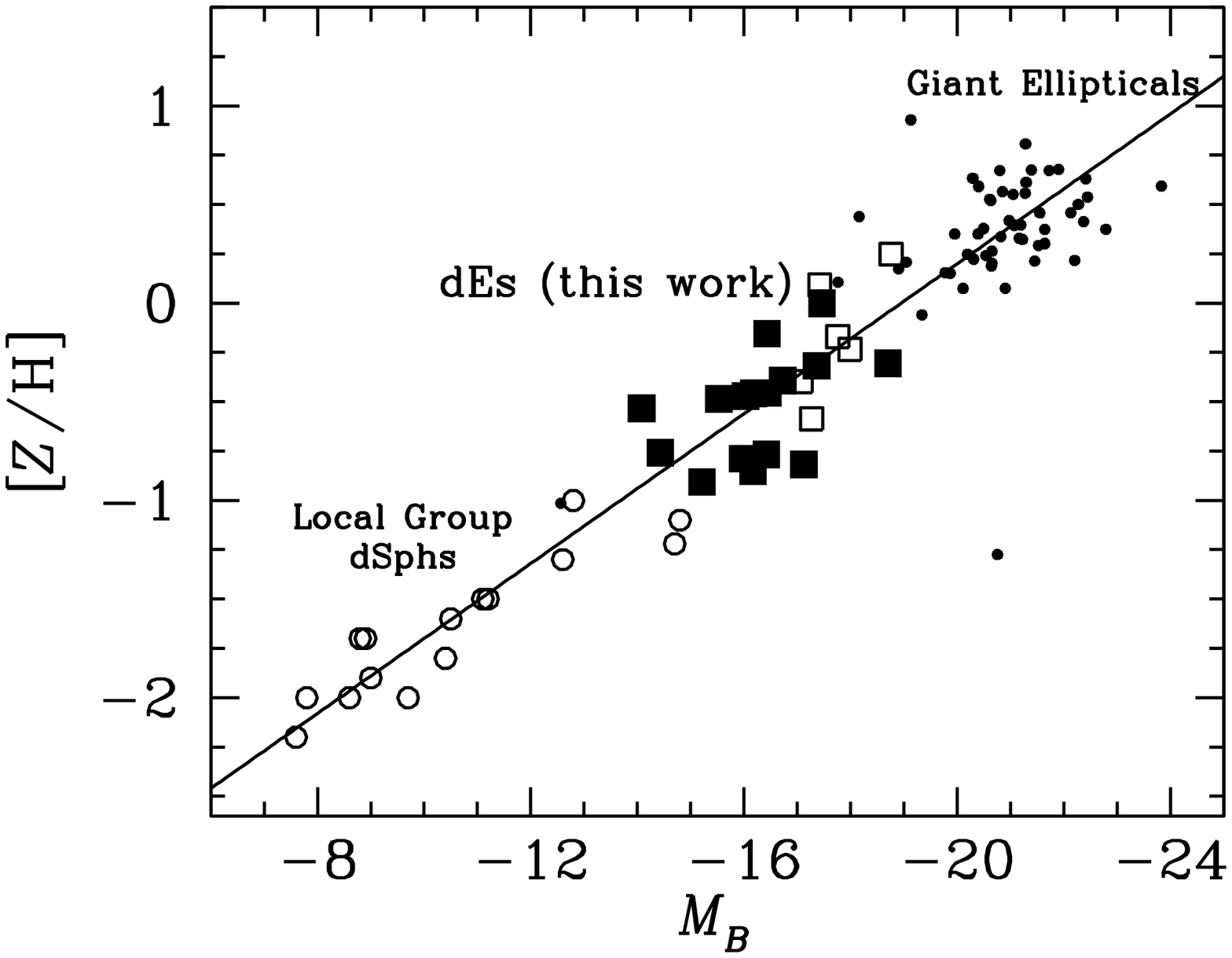}
\includegraphics[width=0.49\textwidth]{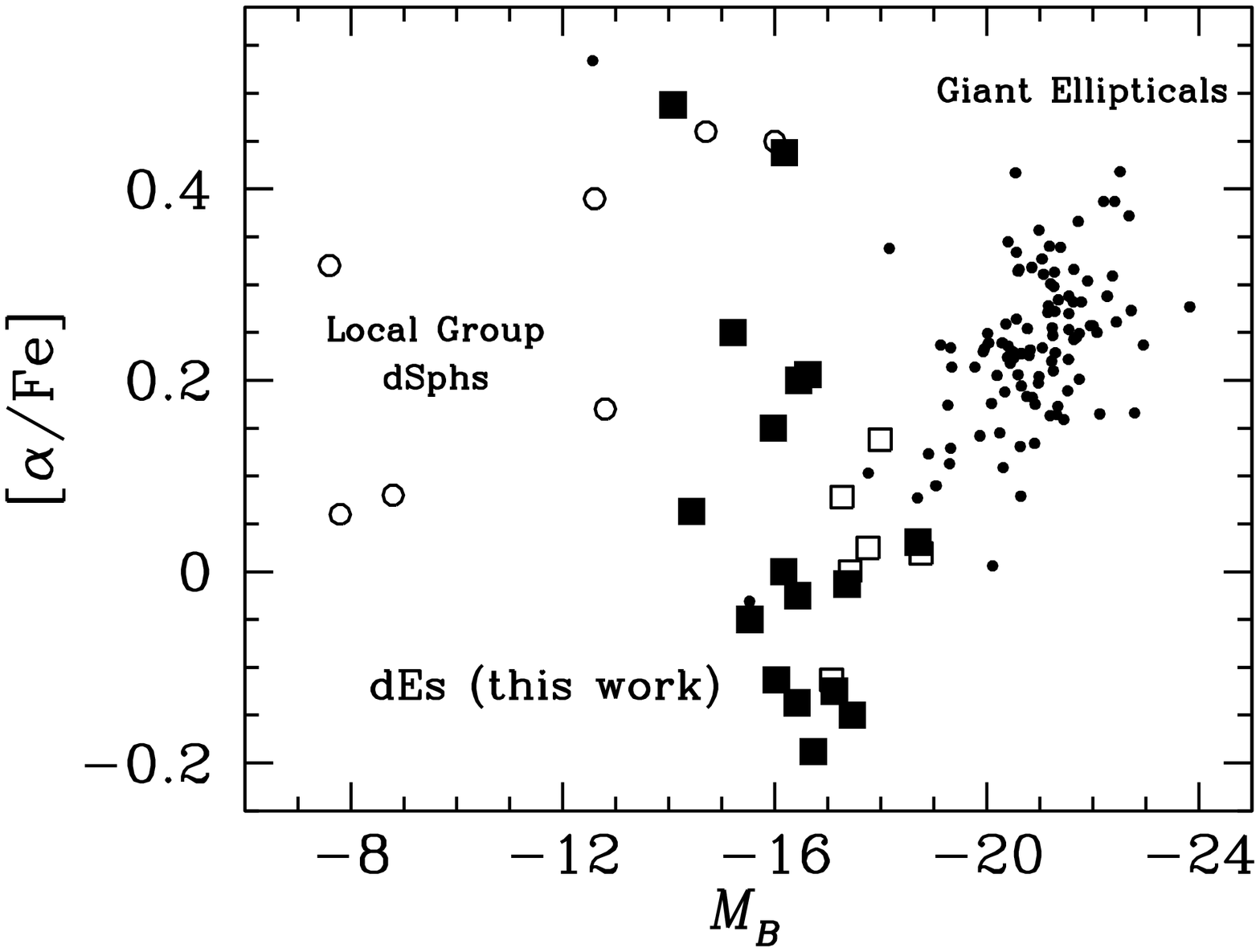}
\end{figure}
In Fig.~2 we plot total metallicity and \aFe\ abundance ratio as
functions of absolute blue luminosity. Our data (filled squares)
covering $-14\geq M_B\geq -19$ nicely fill the gap between the Local
Group dwarf spheroidals and giant elliptical galaxies. We find that
ellipticals follow a surprisingly well defined linear correlation
between absolute magnitude and metallicity over 14 orders of
magnitude. This result suggests that the gas fraction turned into
stars, i.e.\ the efficiency of star formation, which basically
determines the metallicity, steadily increases with increasing mass
and potential well of the object. Hence, the smaller a galaxy, the
larger is the gas fraction it looses through a galactic wind.

The detailed chemical enrichment process, in particular the partition
between Type II and Type Ia supernovae constrained by the \aFe\ ratio,
instead, seems to be very different in dwarf and giant
ellipticals. The dwarf galaxy sample exhibits a large scatter in \aFe,
with a median value of $[\aFe]=0$, which is well below the typical
\aFe\ of giant elliptical galaxies. These relatively low average \aFe\
ratios found here are consistent with the abundance determinations of
individual stars in Local Group dwarf spheroidals (Shetrone et al.\
2001; Tolstoy et al., this volume).  With our larger sample comprising
also fainter objects, we reinforce the conclusion of
\inlinecite{Goretal97}, that also a stellar population dichotomy
exists between dwarf and giant elliptical galaxies. Present-day dwarf
elliptical galaxies are therefore not the fossiled building blocks of
giant ellipticals.

\end{article}
\end{document}